\documentclass[12pt]{article}
\usepackage{geometry}             
\usepackage{graphicx}
\usepackage{amssymb}
\usepackage{amsmath}
\usepackage{epstopdf}

\usepackage[hyperindex=true,
          pdfstartview=FitH,
          bookmarksnumbered=true,
          bookmarksopen=true,
          citecolor=blue,
          linkcolor=blue,
          colorlinks=true,
          pdfborder=001,
          unicode]{hyperref}

\allowdisplaybreaks

\begin{document}

\title{Gravity-mediated holography in fluid dynamics}
\author{Bin Wu and Liu Zhao\\
School of Physics, Nankai University, Tianjin 300071, China\\
{\em email}: \href{mailto:binfen.wu@gmail.com}{binfen.wu@gmail.com} and
\href{mailto:lzhao@nankai.edu.cn}{lzhao@nankai.edu.cn}}
\date{}
\maketitle

\begin{abstract}
For any spherically symmetric black hole spacetime with an ideal fluid source,
we establish a dual fluid system on a hypersurface near the black hole horizon.
The dual fluid is incompressible and obeys Navier-Stokes equation subject to
some external force.  The force term in the fluid equation consists in two parts,
one comes from the curvature of the hypersurface, the other comes from the
stress-energy of the bulk fluid. 
\end{abstract}

\section{Introduction}

Holography is perhaps the most important keyword in the area of high energy theory
over the last twenty years. In spired by the area law of black hole entropy,
't Hooft \cite{t'Hooft} and Susskind \cite{susskind} respectively proposed the 
so-called holographic principle
in the early nineties of the last century. Since then, this concept 
is greatly pushed forward due to Maldacena's discovery of AdS/CFT duality 
\cite{maldcena}.
Holographic dual of both string theory and gravity has been studied extensively,
leading to new developments such as AdS/CMT, AdS/QCD, etc., and applications of
this idea in areas like holographic phase transitions, etc., have now attracted
serious attentions from the condensed matter community.

Generically speaking, holographic duality relates gravity in the bulk to certain
systems with other type of interactions (e.g. gauge interactions) on the boundary.
In some cases the microscopic detail of the boundary degrees of freedom may be
difficult to determine or is unimportant/beyond the scope of concern. In such
circumstances, the boundary degrees of freedom may be coarse grained and analogized
by some fluid component. This is why gravity-fluid duality comes into play. 
The first relationship between gravity and fluid was found quite long ago
\cite{Damor}. In the subsequent 40+ years, similar idea pops up from time to time
until very recently gravity-fluid duality becomes popular and is considered as
a new type of holographic duality \cite{strominger}-\cite{Cai}. In some sense, 
gravity-fluid dual may be 
thought of as a kind of ``poor man's holography''
\cite{1204.2029} because of lack of information on the boundary. 
On the other hand, it also helps to
understand the holographic dual of gravity in general, i.e. not to be limited to
gravity theories with AdS asymptotics.

Although quite a lot of research papers on gravity-fluid duality have been
published, most of the works are on a case by case basis. It is shown in \cite{strominger1} 
that given a particular boost/scaling for Rindler spacetime, there exist a conserved 
stress tensor on boundary which imply the Navier-Stokes equation of the dual fluid,
more relevant works can be found \cite{Naka}-\cite{cai1}, even for higher curvature 
gravity 
case we can also get the same conclusion \cite{Eling1}\cite{BinW}. Another way of 
realizing gravity-fluid duality is to take some special kind of boundary 
conditions to reduce the dynamical degrees of freedom on the boundary of some 
prescribed spacetime background \cite{Strominger2}\cite {xiaoning}. To understand 
gravity-fluid duality more thoroughly, it seems necessary to realize gravity-fluid 
duality on more general footing, for instance, removing either the metric dependence 
or the use of specific boost/scaling technique would greatly increase the usefulness 
of such duality relations. Moreover, it is desirable to consider the holographic 
dual of gravity with matter source, because it has long been known that Einstein 
equation with matter source is equivalent to Navier-Stokes equation for the matter 
source regarded as a fluid \cite{Bhattacharyya}. If one can establish holographic 
dual of sourced gravity in general,
then it is a simple step to establish the duality between the fluid equations of
the matter source and that of the boundary dual. Since this holographic relationship
between the bulk and boundary fluid is established via gravity-fluid duality,
it may be fair to call it as a gravity-mediated holography for the bulk and boundary
fluids. The present work is a first attempt in establishing this kind of
fluid-fluid holography.

\section{Setting up the problem}

Our ultimate aim is to establish a duality relationship between the bulk fluid 
and the boundary fluid. In order to make this idea work, it is necessary to consider
a gravity theory with matter source. We shall deliberately keep the system as
general as possible, though there are still some places where certain kind of
assumptions/restrictions are still needed.

We begin in an $(n+2)$-dimensional spacetime with coordinates $x^\mu =(t,r,
\theta^1, \cdots, \theta^n)$. The Einstein equation with an an ideal fluid source
reads
\begin{align}
&  G_{\mu\nu}=R_{\mu\nu}-\frac{1}{2}Rg_{\mu\nu}=-\Lambda g_{\mu\nu}+T_{\mu\nu},
\label{ein} \\
&  T_{\mu\nu}=(\rho+p)u_{\mu}u_{\nu}+pg_{\mu\nu},
\end{align}
where $\Lambda$ is the cosmological constant, $\rho$, $p$ are respectively the
energy density and pressure in the bulk, $u_{\mu}$ is the proper velocity,
and we have set Einstein's gravitational constant $\kappa=1$.

We assume that there exists a spherically symmetric black hole solution for
eq.(\ref{ein}). In Eddington-Fenkelstein coordinates, the line element
takes the form
\begin{align}
&\mathrm{d}s^2= -f(r)\mathrm{d} t^2+ 2\mathrm{d}t \mathrm{d}r
+r^2 \mathrm{d} \Omega^2, \label{linelem}
\end{align}
where $f(r)$ is a function of $r$ only, $\mathrm{d} \Omega^2$ is the line
element on a unit $n$-sphere which can be written as
\begin{align}
& \mathrm{d} \Omega^2= \mathrm{d} \theta_{1}^2+\sin^2{\theta_{1}}(\mathrm{d}
\theta_{2}^2 +\sin^2{\theta_{2}}(\mathrm{d} \theta_3^2+ \cdots)).
\end{align}
We do not need to solve eq.(\ref{ein}) to get the solution, just the existence of
such a solution is sufficient to facilitate our discussions.

Now consider the timelike hypersurface $\Sigma_c$ determined by taking $r=r_c$
constant. The coordinates on the hypersurface can be naturally taken to be
$x^a=(t,\theta^1,\cdots, \theta^n)$.
The embedding of this hypersurface in the spacetime equips it
with a metric (the first fundamental form) $h_{\mu\nu}$, which is related to the
spacetime metric $g_{\mu\nu}$ via
\begin{align}
& h_{\mu\nu}=g_{\mu\nu}-n_{\mu}n_{\nu}, \label{indmet}
\end{align}
where $n_{\mu}$ is the unit normal vector of $\Sigma_c$. For the spacetime metric
given by the line element (\ref{linelem}), we have
\begin{align}
&n_{\mu}=\left(0,\frac{1}{\sqrt{f}},0,\cdots,0\right), \quad
n^{\mu}=\left(\frac{1}{\sqrt{f}},\sqrt{f},0,\cdots,0\right),\quad n^2=1,
\label{normn}
\end{align}
and the line element corresponding to $h_{\mu\nu}$ reads
\begin{align}
 \mathrm{d} s_{n+1}^2 &= -f(r_c) \mathrm{d} t^2+r_{c}^{2}\mathrm{d} \Omega^2 
 \nonumber \\
 &=-\mathrm{d} x_{0}^2+r_{c}^{2}\mathrm{d} \Omega^2 
 \nonumber \\
 &=-\frac{1}{\lambda^2}\mathrm{d}\tau^2 +r_{c}^{2}\mathrm{d} \Omega^2,
 \label{ndlelm}
\end{align}
where $\tau=\lambda x^{0}=\lambda \sqrt{f}t$ \cite{Strominger2}. The aim for
introducing the rescaling parameter $\lambda$ is to enable us to take the
non-relativistic limit with ease: the limit just corresponds to taking
$\lambda\rightarrow 0$. Since the above line element contains no $dr$ terms, we can
think of the matrix representation of the corresponding metric tensor $h_{\mu\nu}$
as containing a raw and a column full of zero corresponding to the $r$ direction.
Using (\ref{normn}), it is easy to check that $h_{\mu\nu}$ is idempotent, so this
object can also be viewed as the projection operator onto the hypersurface.

The extrinsic curvature (the second fundamental form) of the hypersurface
is defined as
\begin{align}
& K_{\mu\nu}=\frac{1}{2}\pounds_{n}h_{\mu\nu}. \label{K}
\end{align}
The projection of the Einstein equation onto the hypersurface yields some
constraints over the geometric quantities on the hypersurface, and these will be
best expressed using the first and second fundamental forms, as will be
worked out below.

There are two different types of projections, i.e. the longitudinal and
normal projections. These are given respectively as follows:
\begin{align}
& (R_{\mu\nu}-\frac{1}{2}Rg_{\mu\nu})n^{\mu}h^{\nu}{ }_{b}=(-\Lambda g_{\mu\nu}
+T_{\mu\nu})n^{\mu}h^{\nu}{ }_{b},
\end{align}
\begin{align}
(R_{\mu\nu}-\frac{1}{2}Rg_{\mu\nu})n^{\mu}n^{\nu}&=(-\Lambda g_{\mu\nu}
+T_{\mu\nu})n^{\mu}n^{\nu}.
\end{align}
These two equations are usually referred to as momentum and Hamiltonian constraints
respectively.

It will be convenient to make use of the Guass-Coddazi equation to turn the above
constraint equations into a neat form. For this we recall that the following
relations hold,
\begin{align*}
& \hat {R}^{\rho}{ }_{\sigma \mu\nu}=h^{\rho}{ }_{\alpha }h^{\beta }{ }_{\sigma }
h^{\gamma }{ }_{\mu}h^{\delta }{ }_{\nu}R^{\alpha}{ }_{\beta \gamma \delta}
+(K^{\rho}{ }_{\mu}K_{\sigma\nu}-K^{\rho}{ }_{\nu}K_{\sigma\mu}),   \\
& D_{[\mu}K ^{\mu}{ }_{\nu]}=\frac{1}{2}h^{\sigma}{ }_{\nu}R_{\rho\sigma}n^{\rho},
\end{align*}
where $\hat {R}^{\rho}{ }_{\sigma \mu\nu}$ is the curvature tensor on the 
hypersurface and $D_\mu$ is the covariant derivative acting along the hypersurface,
$D_{\mu}h_{\nu\rho}=0$.
Using the relation $h^{\mu}{ }_{\nu}n^{\nu}=0$ and the Guass-Coddazi equation, we
can calculate in a few lines to get the constraint equations on the hypersurface.
The result reads
\begin{align*}
& D_a(K^{a}{ }_{b}-h^{a}{ }_{b}K)=T_{\mu\nu}n^{\mu}h^{\nu}{ }_{b},   \\
& \hat{R}+K^{ab}K_{ab}-K^2=2\Lambda - 2T_{\mu\nu}n^{\mu}n^{\nu}.
\end{align*}
We shall see that these constraints reduce to Navier-Stokes equations on the
hypersurface if appropriate boundary conditions are imposed.

\section{Boundary conditions on the hypersurface}

The boundary conditions we would like to impose on the hypersurface is the so-called
Petrov-like condition \cite{petrov}. To describe such conditions, we need to
project the bulk Weyl tensor onto the hypersurface. Inserting the Einstein equation
(\ref{ein}) into the $(n+2)$-dimensional Weyl tensor
\begin{align*}
& C_{\mu\nu\sigma\rho}=R_{\mu\nu\sigma\rho}
-\frac{2}{n}(g_{\mu [ \sigma}R_{\rho ] \nu}-g_{\nu [ \sigma}R_{\rho ] \mu})
+\frac{2}{n(n+1)}R \,g_{\mu[\sigma}\,g_{\rho]\nu},
\end{align*}
we get the following relation,
\begin{align}
& C_{\mu\nu\sigma\rho}=R_{\mu\nu\sigma\rho}-\frac{4(\Lambda-T)}{n(n+1)}
g_{\mu [ \sigma}g_{\rho ] \nu} -\frac{2}{n}(g_{\mu [ \sigma}T_{\rho ] \nu}
-g_{\nu [ \sigma}T_{\rho ] \mu}),
\end{align}
where $T_{\mu\nu}$  and $T$ are respectively the bulk stress-energy tensor and its
scalar trace. Later on, we shall meet the following projections of the Weyl tensor,
\begin{align*}
& C_{abcd}=h^{\mu}{ }_{a}h^{\nu}{ }_{b}h^{\sigma}{ }_{c}h^{\rho}{ }_{d}
C_{\mu\nu\sigma\rho},   \\
& C_{abc(n)}=h^{\mu}{ }_{a}h^{\nu}{ }_{b}h^{\sigma}{ }_{c}n^{\rho}
C_{\mu\nu\sigma\rho},   \\
& C_{a(n)b(n)}=h^{\mu}{ }_{a}n^{\nu}h^{\sigma}{ }_{b}n^{\rho}
C_{\mu\nu\sigma\rho},
\end{align*}
which can all be re-expressed in terms of the intrinsic and extrinsic curvatures of
the hypersurface and the bulk stress-energy tensor, i.e.
\begin{align}
C_{abcd}&=\hat{R}_{abcd}-K_{ac}K_{bd}+K_{ad}K_{bc}
 -\frac{4(\Lambda-T)}{n(n+1)}h_{a [  c}h_{d ] b}  \nonumber \\
&\quad -\frac{2}{n}h^{\mu}{ }_{a}h^{\nu}{ }_{b}h^{\sigma}{ }_{c}
h^{\rho}{ }_{d}(g_{\mu [ \sigma}T_{\rho] \nu}-g_{\nu [ \sigma}T_{\rho ] \mu}),
\label{C1} \\
C_{abc(n)}&=D_{a}K_{bc}-D_{b}K_{ac}-\frac{2}{n}h^{\mu}{ }_{a}h^{\nu}{ }_{b}
h^{\sigma}{ }_{c}n^{\rho}(g_{\mu [ \sigma}T_{\rho] \nu}-g_{\nu[ \sigma}
T_{\rho] \mu}),   \\
C_{a(n)b(n)}&=-\hat{R}_{ab}+KK_{ab}-K_{ac}K^{c}{ }_{b}
+h^{\mu}{ }_{a}h^{\sigma}{ }_{b}R_{\mu\sigma}
-\frac{2(\Lambda-T)}{n(n+1)}\,h_{ab}   \nonumber \\
&\quad -\frac{2}{n}h^{\mu}{ }_{a}n^{\nu}h^{\sigma}{ }_{b}n^{\rho}
(g_{\mu[ \sigma}T_{\rho] \nu}-g_{\nu[ \sigma}T_{\rho] \mu}). \label{C3}
\end{align}

The boundary conditions to be imposed on the hypersurface $\Sigma_c$
are given by  \cite{petrov}
\begin{align}
&  C_{(l)i(l)j}=l^{\mu}m_{i}{}^{\nu}l^{\sigma}m_{j}{}^{\rho}C_{\mu\nu\sigma\rho}=0,
\label{bdry}
\end{align}
where the Newman-Penrose-like basis vector fields are introduced
which obey
\begin{align}
&  l^2=k^2=0,\,(k,l)=1,\,(l,m_{i})=(k,m_{i})=0,\,(m_{i},m_{j})=\delta^{i}{ }_{j}.
\end{align}
In $(n+2)$-dimensional spacetime, we have $n$ basis vectors $m_i$, $i=1,\cdots,n$. 
$m_i{}^\mu$ is the $\mu$-th component of $m_i$. 
For our specific choice of spacetime metric (\ref{linelem}), the basis vectors
are given as follows \cite{Newman},
\begin{align*}
& m_{i}=\frac{1}{ \sin{\theta_{1}}\cdots\sin{\theta_{i-1}}} \partial_{i},\\
&l=\frac{1}{\sqrt{2}}\left(\frac{1}{\sqrt{f}}\partial_{t}-n\right)
=\frac{1}{\sqrt{2}}\left(\partial_{0}-n\right),\\
&k=-\frac{1}{\sqrt{2}}\left(\frac{1}{\sqrt{f}}\partial_{t}+n\right)
=-\frac{1}{\sqrt{2}}\left(\partial_{0}+n\right).
\end{align*}
Using these we can expand (\ref{bdry}) as
\begin{align}
 & C_{0i0j}+C_{0ij(n)}+C_{0ji(n)}+C_{i(n)j(n)}=0, \label{BBBB}
\end{align}
where, by use of (\ref{C1})-(\ref{C3}), we have
\begin{align}
&  C_{0i0j}=\hat{R}_{0i0j}-K_{00}K_{ij}+K_{0j}K_{0i}-\frac{4(\Lambda-T)}{n(n+1)}
h_{0[  0}h_{j]  i}\nonumber\\
&\qquad\qquad\quad-\frac{2}{n}h^{\mu}{ }_{0}h^{\nu}{ }_{i}
h^{\sigma}{ }_{0}h^{\rho}{ }_{j}(g_{\mu[ \sigma}T_{\rho] \nu}-g_{\nu[ \sigma}
T_{\rho] \mu}),   \\
&  C_{0ij(n)}=D_{0}K_{ij}-D_{i}K_{0j}-\frac{2}{n}h^{\mu}{ }_{0}h^{\nu}{ }_{i}
h^{\sigma}{ }_{j}n^{\rho}(g_{\mu[ \sigma}T_{\rho] \nu}
-g_{\nu[ \sigma}T_{\rho] \mu}),  \\
&  C_{i(n)j(n)}=-\hat{R}_{ij}+KK_{ij}-K_{ik}K^{k}{ }_{j}
+h^{\mu}{ }_{i}h^{\sigma}{ }_{j}R_{\mu\sigma}-\frac{2(\Lambda-T)}{n(n+1)}h_{ij}
\nonumber\\
&\qquad\qquad\quad-\frac{2}{n}h^{\mu}{ }_{i}n^{\nu}h^{\sigma}{ }_{j}
n^{\rho}(g_{\mu[ \sigma}T_{\rho] \nu}-g_{\nu[ \sigma}T_{\rho] \mu}).
\end{align}

In the absence of matter source, such as the case in \cite{Strominger2}, 
the dynamical degrees of freedom on the hypersurface are fully encoded in the
extrinsic curvature $K_{ab}$, or equivalently in the Brown-York tensor
\begin{align}
t_{ab}=h_{ab}K-K_{ab}.   \label{B-Y}
\end{align}
The total number of independent components of
$K_{ab}$ is $\frac{(n+1)(n+2)}{2}$. However, the Petrov-like conditions provides
$\frac{n(n+1)}{2}-1$ constraints over the extrinsic curvature tensor,
where the extra $-1$ is provided by the traceless condition for the Weyl tensor.
Therefore, there are
only $n+2$ remaining degrees of freedom. These remaining degrees of freedom
have to obey the Hamiltonian and momentum constraints described earlier in
Section 2, which will be turned into Navier-Stokes equation for the fluid
on the hypersurface. In this process, these $n+2$ degrees of freedom will be turned
into the density, pressure and velocity components of the boundary fluid.

From the definition(\ref{B-Y}) we get
\begin{align*}
K^{a}{ }_{b}=\frac{1}{n} h^{a}{ }_{b}t-t^{a}{ }_{b},\,\,\,\,t=t^{a}{ }_{a}
=nK^{a}{ }_{a}=nK.
\end{align*}
So,
\begin{align*}
&  K=\frac{t}{n},  \quad\qquad\quad K^{\tau}{ }_{i}=-t^{\tau}{ }_{i},  \\
&  K^{\tau}{ }_{\tau}=\frac{t}{n}-t^{\tau}{ }_{\tau},\quad
K^{i}{ }_{j}=\frac{t}{n}h^{i}{ }_{j}-t^{i}{ }_{j}.
\end{align*}
Inserting these relations as well as (\ref{C1})-(\ref{C3}) into (\ref{BBBB}),
the boundary condition becomes
\begin{align}
 0&=\frac{2}{\lambda^2}t^{\tau}{ }_{i}t^{\tau}{ }_{j}+\frac{t^2}{n^2}h_{ij}-\frac{t}
 {n}t^{\tau}{ }_{\tau}h_{ij}+t^{\tau}{ }_{\tau}t_{ij}+2\lambda\partial_{\tau}
 \left(\frac{t}{n}h_{ij}-t_{ij}\right)-\frac{2}{\lambda}
 D_{{(}i}t^{\tau}{ }_{j{)}}-t_{ik}t^{k}
 { }_{j} \label{b}  \nonumber\\
 &\quad -\frac{1}{n}(T_{\nu\rho}n^{\nu}n^{\rho}-2\Lambda+T+T_{00}-2T_{\rho 0}
 n^{\rho})h_{ij}+T_{ij} - \hat{R}_{ij}.
\end{align}
The calculation in getting this result is somewhat lengthy. We have made the 
coordinate transformation $x^0\to\tau$, resulting in elicit coordinate index 
$\tau$ appearing here and there, e.g. for the second and third terms on the 
right-hand side of (\ref{C1}), we have
\begin{align*}
-K_{00}K_{ij}&+K_{0j}K_{0i}     \\
& =K^{0}{}_{0}K_{ij}+K^{0}{}_{j}K^{0}{}_{i}
=K^{\tau}{}_{\tau}K_{ij}+\frac{1}{\lambda^2}K^{\tau}{}_{i}K^{\tau}{}_{j}     \\
&=\frac{1}{\lambda^2}t^{\tau}{ }_{i}t^{\tau}{ }_{j}
+\frac{t^2}{n^2}h_{ij}-\frac{t}
{n}t^{\tau}{ }_{\tau}h_{ij}+t^{\tau}{ }_{\tau}t_{ij}-\frac{t}{n} t_{ij},
\end{align*}
where raising of indices 0 is made using the metric $h_{\mu\nu}$. We also used 
$\hat{R}_{0i0j}=0$, which is implied by our metric ansatz.

So far, we haven't paid a single word on the boundary conditions to be imposed on
the matter source. In the presence of matter source in the bulk, there are some
extra degrees of freedom for the matter. We also need to impose appropriate
boundary conditions on these extra degrees of freedom. This will not remove all
degrees of freedom for the matter source. The remaining degrees of freedom would
then play some role in the fluid equation on the boundary. Naturally,
the role played by these extra degrees of freedom will be some external impact force
\cite{S Bhattacharyya}\cite{Cai} or fluid source.
In this paper, we will restrict our discussion to spherically symmetric spacetimes
with a perfect fluid source. In this case, the most appropriate boundary condition
to be imposed on the matter source will be fixing $\rho(r)$, $p(r)$ and $u^{\tau}$
to be constant on the hypersurface at $r=r_c$. We will see that such boundary
conditions do work well in establishing the gravity-fluid duality.

\section{Near-horizon limit}

\subsection{Perturbative expansion of the Brown-York tensor}

In this section we shall make perturbative expansions of the Brown-York tensor
and estimate the order of various quantities that appear in the boundary conditions 
and the Hamiltonian and momentum constraints. 

The perturbed Brown-York tensor can be written as \cite{Strominger2}
\begin{align}
t^{a}{ }_{b}=\sum^{\infty}_{n=0} \lambda^{n}t^{a}{ }_{b}^{(n)},
\end{align}
where we have intentionally taken the expansion parameter to be identical to the 
scaling parameter $\lambda$ appeared in (\ref{ndlelm}), so that the perturbative 
limit $\lambda\to 0$ is simultaneously the non-relativistic limit.
In this expansion, the $n=0$ term corresponds to the original (unperturbed) 
background. Using the
definition (\ref{K}), it is straightforward to calculate the components of the
extrinsic curvature, which read
\begin{align*}
& K^{\tau}{ }_{\tau}=\frac{1}{2\sqrt{f}}\partial_{r}f, \quad K^{\tau}{ }_{i}=0,  \\
& K^{i}{ }_{j}=\frac{\sqrt{f}}{r}\delta^{i}{ }_{j},\quad
K=\frac{1}{2\sqrt{f}}\partial_r f+ \frac{n\sqrt{f}}{r}.
\end{align*}
Therefore, to the first order in $\lambda$, we can write the Brown-York stress
tensor as follows,
\begin{align*}
&t^{\tau}{}_{\tau}=\frac{n\sqrt{f}}{r}+\lambda\, t^{\tau}{ }_{\tau}^{(1)}
+ \cdots,   \\
&t^{\tau}{ }_{i}=0+\lambda\, t^{\tau}{ }_{i}^{(1)}  + \cdots, \\
&t^{i}{}_{j}=\left(\frac{1}{2\sqrt{f}}\partial_{r}f
+\frac{(n-1)\sqrt{f}}{r}\right)\delta^{i}{ }_{j}
+\lambda\, t^{i}{ }_{j}^{(1)} + \cdots,   \\
&t=\frac{n}{2\sqrt{f}}\partial_{r}{f}+ \frac{n^2 \sqrt{f}}{r}
+ \lambda\, t^{(1)}+ \cdots,
\end{align*}
where $t = t^a{}_a$ is the trace of the Brown-York tensor.
We assume that the smooth function $f(r)$ has some zeros at $r=r_h$, which
corresponds to the event horizon of a black hole. In this case, we can expand $f(r)$
near the horizon, i.e.
\begin{align*}
f(r_c)&= f^{\prime}(r_{h})(r_c-r_h)+ \frac{1}{2}f^{\prime \prime}(r_h)(r_c-r_h)^2
+\cdots,
\end{align*}
where $r_c$ is the fixed constant determining the hypersurface $\Sigma_c$. For
the above expansion to make sense, we require the hypersurface to be
located near the horizon, so that $r_c-r_h$ is small. In particular, 
we let this difference to be related to $\lambda$ via $r_c-r_{h}= \alpha^2 
\lambda^2$, where $\alpha$ is an arbitrary fixed constant which is introduced to 
balance the dimensionality. Then, it is easy to see that provided 
$f'(r_h)\ne 0$,
\begin{align}
&f \rightarrow \mathcal{O}(\lambda^2)+ \mathcal{O}(\lambda^4)+\cdots, \nonumber  \\
& \frac{f}{r} \rightarrow \mathcal{O}(\lambda^2)+ \mathcal{O}(\lambda^4)+\cdots,
\nonumber  \\
& \sqrt{f}=\sqrt{(r_c-r_h)}\sqrt{f^{\prime}(r_{h})+\frac{1}{2}f^{\prime \prime}(r_h)
(r_c-r_h)+\cdots} \rightarrow \mathcal{O}(\lambda)+\mathcal{O}(\lambda^3)+\cdots,
\nonumber \\
& \frac{\sqrt{f}}{r} \rightarrow \mathcal{O}(\lambda)+\mathcal{O}(\lambda^3)+\cdots,
\nonumber \\
& \partial_{r}f= f^{\prime}(r_{h})+ f^{\prime \prime}(r_h)(r_c-r_h) 
+\cdots \rightarrow \mathcal{O}(\lambda^{0})
+\mathcal{O}(\lambda^2)+\cdots,    \nonumber \\
& \frac{\partial_r f}{\sqrt{f}} \rightarrow \mathcal{O} (\lambda^{-1})
+ \mathcal{O}(\lambda) + \cdots.  \label{c}
\end{align}

\subsection{Matter perturbations}

In the bulk, we take the matter source to be an ideal fluid,
\begin{align}
T_{\mu\nu}=(\rho+p)u_{\mu}u_{\nu}+pg_{\mu\nu},
\end{align}
where $\rho=\rho(r)$, $p=p(r)$ because of spherical symmetry, and
$u^{\mu}=(-\frac{1}{\sqrt{f}},0,\cdots,0)$ is the proper velocity 
satisfying $u^2=-1$.
Recall that the bulk matter contributes in the boundary condition via (\ref{b}).
We can evaluate explicitly such contribution by calculating the relevant terms
in (\ref{b}):
\begin{align}
&  -\frac{1}{n}(T_{\nu\rho}n^{\nu}n^{\rho}-2\Lambda+T+T_{00}-2T_{\rho 0}
n^{\rho})h_{ij}+T_{ij}  \nonumber \\
& \qquad=-\frac{1}{n}(-2\Lambda+2p+(f-1)\rho)h_{ij}.
\end{align}
Since we have taken $\rho,p$ and $u^{\tau}$ to be fixed on the hypersurface,
the only possible perturbation on the bulk matter must be realized via perturbing
the other components of the proper velocity, i.e.
\begin{align}
&  u^{r}=0+\lambda u ^{r(1)}+ \cdots,      \nonumber \\
&  u^{i}=0+ \lambda u^{i(1)}+ \cdots.
\end{align}
Using $u_{\mu}=g_{\mu\nu}u^{\nu}$, we can get
\begin{align}
u_{\mu}=\left(\sqrt{f}+\lambda u ^{r(1)}+ \cdots, \,  -\frac{1}{\sqrt{f}}, \,
h_{ij}(\lambda u^{i(1)}+ \cdots)\right).
\end{align}
In the next section, these perturbative expansions will be inserted into the
perturbed version of the Hamiltonian and momentum constraints to yield the correct
fluid equations on the hypersurface.

\section{Fluid on the hypersurface}

Now we are in a position to combine all previous analysis to write down the fluid
equations on the hypersurface. To do so, we first consider the perturbation of the
boundary condition (\ref{b}). With the aid of the order estimation (\ref{c}), we get
in the first nontrivial order $\mathcal{O}(\lambda^{(0)})$ the following relation,
\begin{align}
 \frac{\sqrt{f'_h}}{\alpha}t^{i}{ }_{j}^{(1)}&= 2t^{\tau}{ }_{k}^{(1)}t^{\tau}{ }_{j}^{(1)}h^{ik(0)}-2h^{ik(0)}D_{(j}
 t^{\tau}{ }_{k)}^{(1)}+ \frac{\sqrt{f'_h}}{n \alpha} t^{(1)} h^{i}{ }_{j}^{(0)}\nonumber\\
 &\qquad -\frac{1}{n}\left(-2\Lambda+2p-\rho+\frac{n \partial_r f}{r}|_{r_h}
 \right)h^{i}{ }_{j}^{(0)} -\hat{R}^{i}{ }_{j},    \label{d}
\end{align}
where $f'_h$ represent the derivative of function $f$ evaluate at $r_h$ and $h_{ij}^{(0)}$ is the metric of a standard $n$-sphere of radius $r_h$,
$h_{ij}^{(0)}=r_{h}^2 d \Omega^2_{n}$.

Now, consider the momentum constraints on hypersurface. Inserting the definition of Brown-York stress tensor, we get
\begin{align}
D_{a}t^{a}{ }_{b}=-T_{\mu b} n^{\mu}. \label{divnf}
\end{align}
Let us expand this equation in components. First look at the $\tau$ component.
The left-hand side of this component reads
\begin{align}
 D_{a}t^{a}{ }_{\tau}&=D_{\tau} t^{\tau}{ }_{\tau}+D_{i} t^{i}{ }_{\tau}
 \nonumber \\
 &= D_{\tau} t^{\tau}{ }_{\tau} - \frac{1}{\lambda^2}D_{i}(t^{\tau}{ }_{j}h^{ij}).
 \label{taucomp}
\end{align}
We can easily make some order estimation for each term appearing on the 
right-hand side,
\begin{align}
&  D_{\tau} t^{\tau}{ }_{\tau} = D_{\tau} t^{\tau}{ }_{\tau}^{(1)}+\cdots
\rightarrow \mathcal{O}(\lambda^{(1)}),  \nonumber \\
&  \frac{1}{\lambda^2}D_{i}(t^{\tau}{ }_{j}h^{ij})= \frac{1}{\lambda} h^{ij(0)}D_{i}
t^{\tau}{ }_{j}^{(1)}+\cdots \rightarrow \mathcal{O} (\lambda^{(-1)}).
\end{align}
Therefore, we determine that the leading order of (\ref{taucomp}) is
$\frac{1}{\lambda} h^{ij(0)}D_{i}t^{\tau}{ }_{j}^{(1)} $.
The right-hand side of the $\tau$ component of (\ref{divnf}) is
\begin{align}
T_{\mu \tau}n^{\mu}= T_{\tau \tau}n^{\tau}+T_{r\tau}n^{r}
=(\rho+p)(u_{\mu}n^{\mu})u_{\tau}\rightarrow \mathcal{O}(\lambda^{(1)})+\cdots.
\end{align}
Since the leading order of $(\rho+p)$ is $\mathcal{O}(\lambda^{(0)})$, $u_{\mu}
n^{\mu}=\frac{\lambda}{\sqrt{f}}u^{(1)}_{r}+\cdots$ is $\mathcal{O}(\lambda^{(0)})$,
and $u_{\tau}=\sqrt{f}+\,\lambda u ^{r(1)}+ \cdots$ is
$\mathcal{O} (\lambda^{(1)})$,  the entire term is $\mathcal{O} (\lambda^{(1)})$.
So, we get the first useful equation at $\mathcal{O}(\lambda^{(-1)})$:
\begin{align}
D_{i}t^{\tau i(1)}=0. \label{divfree}
\end{align}

Next we consider the spacial components of the momentum constraints.
We have\begin{align}
& D_{a}t^{a}{ }_{i}=D_{\tau}t^{\tau}{ }_{i}+D_{j}t^{j}{ }_{i}. \label{Dt}
\end{align}
Inserting (\ref{d}) into the above equation, we get in the order
$\mathcal{O}(\lambda^{(1)})$ the following expression,
\begin{align}
& D_{\tau}t^{\tau}{ }_{i}^{(1)}+\frac{\alpha}{\sqrt{f'_h}}\left(2t^{\tau}{ }_{i}^{(1)}
D^{k}t^{\tau}{}_{k}^{(1)}+2t^{\tau j (1)}D_{j}t^{\tau}{ }_{i}^{(1)}-D^{k}
(D_{i}t^{\tau}{ }_{k}^{(1)}+D_{k}t^{\tau}{ }_{i}^{(1)})
+\frac{\sqrt{f'_h}}{n \alpha}D_{k}t^{(1)}-D_{j}\hat {R}^{j}{ }_{i}\right).    \label{e}
\end{align}
It can be shown that the components $\hat\Gamma^{a}_{\tau i}$ of the Christoffel
symbol on the hypersurface $\Sigma_c$ are vanishing, we have
\begin{align}
& D_{\tau}t^{\tau}{ }_{i}^{(1)}=\partial_{\tau}t^{\tau}{ }_{i}^{(1)}
-\hat\Gamma^{a}{ }_{\tau i}t^{\tau}{ }_{a}^{(1)}
=\partial_{\tau}t^{\tau}{ }_{i}^{(1)}.
\end{align}
This gives a simplification to the first term in (\ref{e}). The second term in (\ref{e}) is zero thanks to (\ref{divfree}). Using the same
equation (\ref{divfree}), we can also get
\begin{align}
D_{k}D_{i}t^{\tau k (1)}=(D_{k}D_{i}-D_{i}D_{k})t^{\tau k(1)}=\hat{R}_{im}t^{\tau m (1)}.
\end{align}
This quantity has appeared in the 4-th term in (\ref{e}). Now,
according to Bianchi identity, we have
\[
D_{a}\hat{R}^{a}{ }_{i}=  \frac{1}{2}D_{i}\hat{R}.
\]
Our specific choice of line element on the hypersurface implies that
$\hat R^{\tau}{}_i$ is vanishing and $\hat R$ is constant. So, we have
\begin{align}
D_{a}\hat{R}^{a}{ }_{i}=D_{j}\hat{R}^{j}{ }_{i}=  \frac{1}{2} D_{j}\hat{R}=0.
\end{align}
So, the last term in (\ref{e}) is also vanishing. In the end, (\ref{Dt}) becomes
\begin{align}
 D_{a}t^{a}{ }_{i}
=\partial_{\tau}t^{\tau}{ }_{i}^{(1)}
+\frac{\alpha}{\sqrt{f'_h}}\left(2t^{\tau j(1)}D_{j}t^{\tau}{ }_{i}^{(1)}
-D_{k}D^{k}t^{\tau}{ }_{i}^{(1)}-\hat R^{m}{ }_{i}t^{\tau}{ }_{m}^{(1)}\right)
+\frac{1}{n}D_{k}t^{(1)} \label{EL}
\end{align}
in the leading order $\mathcal{O} (\lambda^{(1)})$.

Next, let us consider the contribution of bulk matter to the momentum constraint.
Evidently,
\begin{align*}
T_{\mu i}n^{\mu}&=T_{\tau i}n^{\tau}+T_{r i}n^{r} =(\rho+p)(u_{\mu}n^{\mu})u_{i}.
\end{align*}
The leading order of this term is $\mathcal{O}(\lambda^{(1)})$,
\begin{align}
T_{\mu i}n^{\mu} =\lambda (p+\rho)(u_{\mu}n^{\mu})^{(0)}h^{(0)}{ }_{ij}u^{j(1)}
+ \mathcal{O}(\lambda^{(2)}). \label{ER}
\end{align}

It is crucial to introduce the following new notations,
\begin{align}
t^{\tau}{ }_{i}^{(1)}=\frac{v_{i}}{2},\quad \frac{t^{(1)}}{n}=\frac{\hat{p}}{2}, 
\quad f_{i}=-T_{\mu i}n^{\mu(1)}. \label{notatio}
\end{align}
Here $v_{i} $ and $\hat{p}$ are to be interpreted respectively as the velocity and 
pressure of the fluid on the hypersurface. Inserting these new notations into
(\ref{EL}) and (\ref{ER}) and then substituting the results into
(\ref{divnf}), we get the following equation,
\begin{align}
\partial_{\tau}v_{i}+D_{k}p+ \frac{\alpha}{\sqrt{f'_h}}\left(v^{j}D_{j}v_{i}-D_{k}
D^{k}v_{i}-\hat R^{m}{ }_{i}v_{m}\right)=f_{i}. \label{NS}
\end{align}
Recall that $\alpha$ is an arbitrary finite constant which is introduced to balance 
the dimensionality in the relation $r_c-r_h=\alpha^2 \lambda^2$. We can freely 
choose $\alpha=\sqrt{f'_h}$ as did in \cite{Ying2}. Doing so (\ref{NS}) becomes
the standard forced Navier-Stokes equation in curved background, i.e.
\begin{align*}
\partial_{\tau}v_{i}+D_{k}p+ v^{j}D_{j}v_{i}-D_{k}
D^{k}v_{i}=f_{i}+\hat R^{m}{ }_{i}v_{m}. 
\end{align*}
We intentionally moved the curvature term to the right-hand side, because 
both $f_i$ and the curvature term provide external forces to the fluid equation.  
The force $f_i$ is related to the bulk stress-energy tensor explicitly 
via (\ref{notatio}), while the curvature force is also related to the bulk 
stress-energy tensor implicitly through the bulk Einstein equation.
The equation (\ref{divfree}) is re-interpreted as the incompressible condition
\begin{align}
D_{k}v^{k}=0
\end{align}
of the boundary fluid.

While completing the draft of this paper, a new paper on gravity-fluid
duality appeared on arXiv \cite{WLTZ}, which contains some overlap in interests
with our work. The difference lies in that, in \cite{WLTZ}, gravity-fluid duality
for general non-rotating black holes are established, while in our work only 
spherically symmetric black holes are considered. However, our work still stands in
that we have included an ideal fluid source in the bulk, while in \cite{WLTZ}, only
sourceless gravity is considered. Both works aim to extend gravity-fluid duality 
to more general setting. One can expect that more fruitful result in this field
is upcoming.

\providecommand{\href}[2]{#2}\begingroup
\footnotesize\itemsep=0pt
\providecommand{\eprint}[2][]{\href{http://arxiv.org/abs/#2}{arXiv:#2}}

\end{document}